\documentclass[twocolumn,aps,showpacs]{revtex4}
\usepackage{amsmath}

\usepackage{graphicx}
\usepackage{dcolumn}
\usepackage{bm}

\usepackage{color} 

\begin{document}

\title{Spatiotemporal Amplitude and Phase Retrieval of Bessel-X pulses using a Hartmann-Shack Sensor.}

 \author{F. Bonaretti$^{1}$, D. Faccio$^{1,2}$,  M. Clerici$^{1}$, J. Biegert$^{2,3}$, P. Di Trapani$^{1,4}$}

 \address{$^{1}$CNISM and Department of Physics and Mathematics, University of Insubria, Via Valleggio 11, IT-22100 Como, Italy\\
$^{2}$ICFO-Institut de Ci\`ences Fot\`oniques, Mediterranean Technology Park, 08860 Castelldefels, Barcelona, Spain\\
 $^{3}$ICREA Instituci\'o Catalana de Recerca i Estudis Avan\c cats, 08010 Barcelona, Spain\\
$^{4}$Department of Quantum Electronics, Vilnius University, Sauletekio Ave. 9, bldg. 3, LT-10222, Vilnius, Lithuania}

  \email{daniele.faccio@uninsubria.it}

\date{\today}

\begin{abstract}
We propose a new experimental technique, which allows for a complete characterization of ultrashort optical pulses both in space and in time. Combining the well-known Frequency-Resolved-Optical-Gating technique for the retrieval of the temporal profile of the pulse with a measurement of the  near-field made with an Hartmann-Shack sensor, we are able to  retrieve the spatiotemporal amplitude and phase profile of a Bessel-X pulse. By following the pulse evolution along the propagation direction we highlight the superluminal propagation of the pulse peak.  
\end{abstract}

%
%
\maketitle
%

%

The growth of the applications of ultrashort laser pulses in the last years, has required the development of new techniques to measure them precisely.
Many methods have been studied in order to obtain amplitude and phase information separately in time and space domains. Although such characterization can be sufficient for a large number of applications, a full space-time knowledge of the pulse is crucial in the cases where the use of complicated pulses is necessary.\\ 
\indent Several techniques have been developed to obtain spatial information about monochromatic pulses.
A very well known example of a single-shot method is the Hartmann-Shack (H-S) sensor \cite{Shack:1971}. The H-S sensor is an array of lenslets each with the same focal length. The measurement of the pulse in the focal plane of the sensor allows through a simple algorithm to retrieve the amplitude and phase of the pulse. 
Recently the H-S method has also been extended to polychromatic pulses \cite{Biegert:2005}.\\
\indent In the last decade many efforts have also been made to develop techniques able to fully characterize ultrashort pulses in the time domain. The most wide-spread and well-known is Frequency Resolved Optical Gating (FROG) \cite{Trebino:1993}. This is based on the measurement of a temporally resolved spectrum (spectrogram) and the electrical field is obtained through an iterative inversion algorithm. \\
Techniques have been also developed for characterizing space-time coupled pulses, i.e. pulses whose temporal profile depends on the transverse spatial position.
Among these, Spatial Encoded Arrangement for Temporal Analysis by Dispersing a Pair of Light E-fields (SEA TADPOLE) \cite{Trebino:2006,Trebino:2007}, Spatially and Temporally Resolved Intensity and Phase Evaluation Device: Full Information from a Single Hologram (STRIPED FISH) \cite{Stripedfish:2006} and Spatial Encoded Arrangement SPIDER (SEA SPIDER) \cite{Kosik:2005} and, more recently, complete retrieval of the optical amplitude and phase using the $(k_{\bot},\omega)$ spectrum (CROAK) \cite{Bragheri:2008}. \\
\indent In this letter we propose a simple and economic measurement technique, Shackled-FROG, that retrieves the full spatial and temporal amplitude and phase of ultrashort laser pulses with a very cost-effective and simple layout that may be made in single-shot. The temporal profile of the pulse at a single radial position is characterized by a standard Second Harmonic Generation (SHG) FROG measurement, while the spatial characterization is performed with a home-made H-S sensor. We demonstrate the method by characterizing the propagation of a Bessel-X pulse and the superluminal motion of the central peak is clearly visualized. We conclude with considerations regarding the extension of the method to account for pulses with complex space-time coupled features such as spatial chirp and angular dispersion. \\
\begin{figure}[t]
\centering\includegraphics[width=8cm]{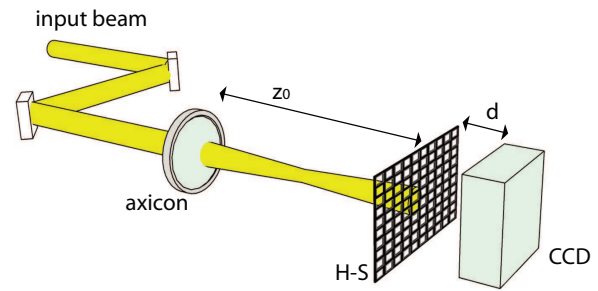}
\caption{\label{fig:setup_icfo} Experimental layout of the experiment: $z_{0}$ indicates the distance between the axicon and the H-S sensor, $d$ indicates the distance between the H-S sensor and the detector. The temporal characterization is performed by substituting the H-S sensor with a FROG setup.}
\end{figure}
\begin{figure}[t]
\centering\includegraphics[width=8cm]{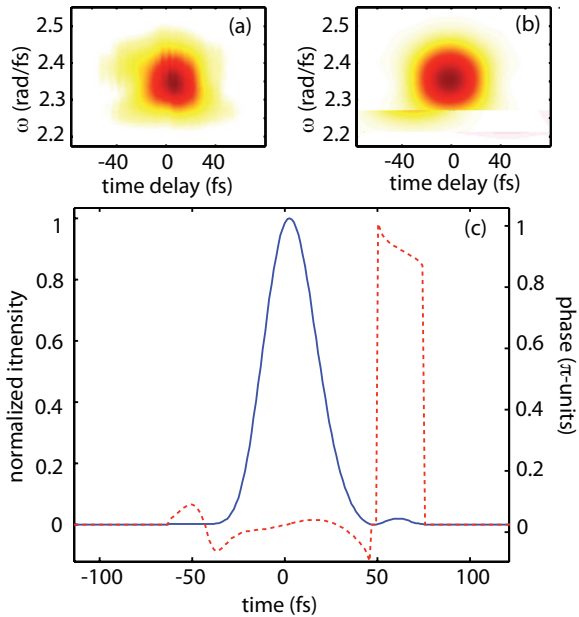}
\caption{\label{fig:frog} (a) Experimental spectrogram. (b) Last spectrogram retrieved by the iterative algorithm. (c) Retrieved temporal pulse: intensity (continuous line) and phase (dashed line). The retrieved pulse duration is $33$ fs FWHM.}
\end{figure} 
\indent The H-S sensor was built following Ref.~\cite{Migdal:2008}: it consists of a regular array of transmissive Fresnel zone plates, that play the same role of the lenses in the standard H-S sensor. The Fresnel zone plate consists of a set of concentric rings, alternatively transparent and opaque. The displacement $\Delta x,y$ of the focused spots from each lenslet with respect to the projection of the centre of the respective Fresnel zone plate on the detector, will give the spatial phase derivative of the pulse through the relation
$\partial z/\partial (x,y)=-\Delta (x,y)/f$, where $z$ is the relative displacement of the wavefront with respect to a flat wavefront and $f$ is the focal length of the single Fresnel lens or the distance between the lens array and the CDD plane on which the H-S pattern is recorded. The position of the displaced focused spots can be accurately computed by using the center of mass of the focusing spots measured.
 By a simple integration the wavefront in terms of $z$ is retrieved and for the adimensional phase it is sufficient to multiply $z$ by $2\pi/\lambda$. This kind of sensor can reach an accuracy of the order of $\lambda$ in absolute measurements \cite{Migdal:2008}.
We note that if the pulse to be measured has a large spectral bandwith or a strong spatial chirp, separate measurements with interference filters (as in Ref.~\cite{Biegert:2005}) or a single measurement with the H-S sensor placed after an imaging spectrometer may be performed. This allows the retrieval of the spatial phase for each wavelength. \\
\indent The spatial phase at each wavelength is uncorrelated with respect to the others. However the link along the temporal axis between each spatial phase map is performed by means of a single FROG measurement at a certain coordinate x,y (e.g. $x,y=0$), albeit with the constraint that all frequencies are present at this spatial position. The full space-time structure of the pulse is thus obtained.\\
\begin{figure}[!t]
\centering\includegraphics[width=8cm]{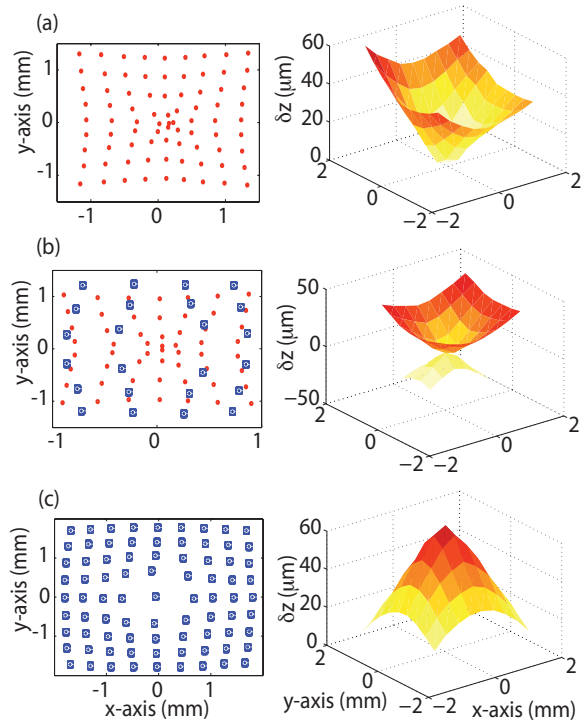}
\caption{\label{fig:fasi} Left hand side figures: H-S acquisition. Right hand side: corresponding retrieved phase. (a) Beginning of the Bessel zone: $d=18$ mm. (b) Middle of the Bessel zone: $d=58$ mm. (c) End of the Bessel zone: $d=108$ mm. In all figures, the converging front is shown in red circles and the diverging pulse front in blue spots.}
\end{figure} 
\indent We applied our technique to characterize a Bessel-X pulse generated by means of an axicon \cite{Saari:1997}. 
We note that in such a pulse all the wavelengths are deflected at the same angle so that there is negligible angular dispersion, i.e. $\theta\neq\theta(\lambda)$, under the condition that the dispersion of the material is negligible. This requirement is fullfilled for example by taking a fused silica axicon with a base angle of $\alpha=2.5^{\circ}$, such as the one used in our experiment (see below), and a $790$ nm pulse with $28$ nm FWHM bandwidth. Indeed the corresponding cone angle variation due to material dispersion between the extremal wavelengths (FWHM) is $\approx 0.0035^{\circ}$ and may be neglected.\\
\indent The experimental layout is shown in Fig. \ref{fig:setup_icfo}. A $790$ nm, $30$ fs, $4$ mm FWHM diameter pulse, delivered by at 1 kHz repetition rate (KM-Labs), impinges on a fused silica axicon with nominal base angle $\alpha=2.5^{\circ}$. The spatial profile of the resulting Bessel-X pulse, whose corresponding Bessel zone length is $\approx 10$ cm, is then measured by placing the H-S sensor at a distance $z_{0}$ after the exit facet of the axicon. The detecting device, a CCD, is placed after the H-S sensor at a distance $d=10$ mm corresponding to the focal length of an H-S sensor designed for the central wavelength of the pulse. The temporal profile of the pulse was measured independently by using standard SHG FROG technique, assuming that no changes occur in the temporal characteristics of the pulse along the direction of propagation. The FROG measurement is shown in Fig. \ref{fig:frog}, we retrieved a pulse with a duration of $33$ fs, with an almost flat phase.  We thus characterize the space-time evolution of the pulse, by scanning the Bessel zone with the H-S sensor. The measurements were repeated at 1 cm intervals over a 10 cm range that covered the whole Bessel zone. Fig. \ref{fig:fasi} shows three significant spatial phase measurements at different distances from the exit facet of the axicon: at the beginning of the Bessel zone (a), in the middle of the Bessel zone (b), and at its end (c). The pattern formed by the spots of each measurement reflects the conical phase shape of the Bessel-X pulse.
We remark that in Fig.~\ref{fig:fasi}(b) the H-S acquisition presents simultaneously two distinct patterns, corresponding to the diverging front of the pulse (squares) and the converging tail (circles) of the pulse. We were able to distinguish between the two contributions, by subtracting the converging pattern in Fig.~\ref{fig:fasi}(a) from Fig.~\ref{fig:fasi}(b) which resulted in the pattern corresponding to the diverging front of the Bessel-X pulse. 
From the retrieved spatial phases, the axicon angle $\alpha$ was found to be $3.10^{\circ}\pm0.14^{\circ}$. Since this value is not the nominal one given by the constructor we performed an independent measurement of the axicon angle $\alpha$, by recording the intensity Bessel profile of the beam. By fitting the zeros of the Bessel profile we found $\alpha=3.16^{\circ}\pm0.10^{\circ}$, in close agreement with the values obtained from the H-S measurements.\\
\begin{figure}[t]
\centering\includegraphics[width=8cm]{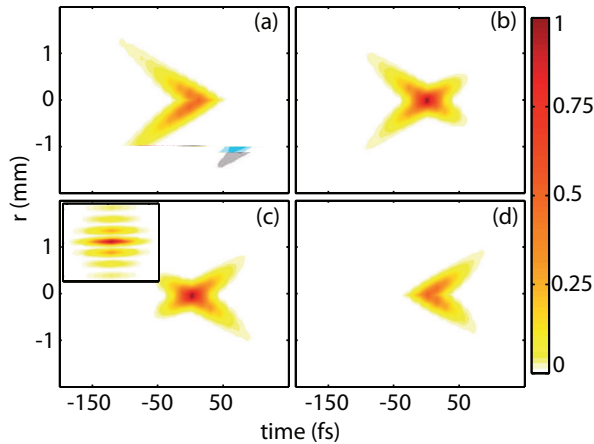}
\caption{\label{fig:evoluzione} Space-time evolution of the Bessel-X pulse during the propagation. The pulses are shown in the reference frame of the peak of the Bessel-X pulse. (a) $d=18$ mm. (b) $d=48$ mm. (c) $d=78$ mm. (d) $d=108$ mm. The inset in (c) shows the Bessel profile formed in the central zone of the pulse.}
\end{figure} 
\indent Linking the temporal and spatial measurements we obtain the spatiotemporal evolution of the Bessel-X pulse at different $z_{0}-$positions within the Bessel zone. Figs.~\ref{fig:evoluzione} (a), (b), (c), and (d) show the space-time evolution of the pulse respectively at $z_{0}=18$ mm (beginning of the Bessel zone), at $z_{0}=48$ mm and $z=78$ mm (in the middle of the Bessel zone), and at $z_{0}=108$ mm (end of the Bessel zone). In Fig. (a) we see only the incoming plane waves. In Fig. (b) and (c) the Bessel-X pulse is formed and the unbalancing between the front of the pulse and its tail is clearly visible. Finally Fig. (d) shows the outgoing plane waves. The inset in Fig. (c) shows a high-resolution enlargement of the center of the corresponding Bessel-X pulse with the expected Bessel profile in the spatial dimension.\\
The temporal reference frame in Fig.~\ref{fig:evoluzione} is centered on the Bessel-X peak and we may clearly observe how the constituent off-axis waves move backwards (toward longer times) for increasing $z$. This is due to the fact that the Bessel-X peak velocity is given by $v_p=v_g/\cos(\theta)$ whilst the center of mass of the overall pulse energy  travels at $v_E=v_g\times\cos(\theta)$, where $v_g=1/(dk/d\omega)$ and the Bessel-X angle , $\theta=[\sin^{-1}(n_{\rm axicon}\sin\alpha)-\alpha]$ depends on the axicon refractive index, $n_{\rm axicon}=1.45$. We may therefore predict a Bessel peak velocity $v_p=1.000296\times c_0$, where $c_0$ is the vacuum velocity of light. By retrieving the differences in the Bessel peak and the energy center-of-mass positions between successive propagation distances we retrieve an experimental value of $v_p=1.00022\pm0.00008\times c_0$. This is in close agreement with the predicted value and provides a clear and simple measurement of the superluminality of the Bessel-X peak \cite{mugnai,milchberg,recami}.\\
\indent In conclusion, we propose a new technique based on the combination of FROG, and Hartmann-Shack near-field measurements to completely characterize in space and time the amplitude and phase of an ultrashort pulse. The technique gives the full (x,y,t) amplitude and phase of the laser pulse and may be made single shot if the FROG measurement is substituted by a Grenouille or SPIDER measurement. The Bessel-X pulse is a very simple example of a space-time coupled pulse and does not present angular dispersion or spatial chirp in the conditions adopted in this work. The present technique may be extended to include these couplings by use of an imaging spectrometer so that the H-S sensor is used to measure the phase of the $r-\lambda$ spectrum. Future work will be directed at implementing this extension and to the measurement of more complex pulses.

\indent The authors wish to acknowledge  support from the Consorzio Nazionale Inter-unversitario per le Scienze della Materia (CNISM), progetto INNESCO.  PDT acknowledges support from Marie Curie Chair project STELLA, Contract No. MEXC-CT-2005-025710. D.F. acknowledges support from Marie Curie grant, Contract No. PIEF-GA-2008-220085. JB acknowledges support from the Spanish Ministry of Education and  Science through its Consolider Program Science (SAUUL - CSD 2007-00013) as well as  through ÒPlan NacionalÓ (FIS2008-06368-C02-01). MC acknowledges support MIUR, project RBIN04NYLH.

\end{document}